\documentclass[10pt]{article}
\title{Lindbladian-Induced Alignment in Quantum Measurements}
\author{R. Englman and A. Yahalom\\ Ariel University, Ariel 40700,Israel}

\usepackage{graphicx}
\begin{document}
\maketitle

\newcommand{\beq}{\begin{equation}}
\newcommand{\enq}{\end{equation}}
\newcommand{\ber}{\begin{eqnarray}}
\newcommand{\enr}{\end{eqnarray}}


~

\begin{abstract}
~

An expression of the Lindbladian form is proposed that ensures an unambiguous  time-continuous reduction of the initial system-pointer wave-packet to  one  in which the readings and the observable's values are aligned, formalized as the transition from an outer product to an inner product of the system's and apparatus' density matrices. The jump operators are in the basis of the observables, with uniquely determined parameters derived from the measurement set-up (thereby differing from S. Weinberg's Lindbladian resolution of wave-packet formalism) and conforming to Born's probability rules. The novelty lies in formalising the adaptability  of the surroundings (including the measuring device) to the mode of observation. Accordingly, the transition is of finite duration (in contrast to its instantaneousness in the von Neumann's formulation). This duration  is estimated for a simple half-spin-like model.
\end {abstract}

\section{Introduction}\label{introduction}
In the century-run of quantum physics (\textit{plus} 4 years, if one marks its beginning with the award of  a Nobel Prize in 1918 to Max Planck for "his discovery of quanta") a single  shadow of \textit{non-sequitur}  has darkened its glorious achievements, one that goes variously under the names  of wave-function collapse, reduction of the wave-packet, quantum measurement, einselection, etc. Aspects of the problem (or its articulations) were manifold, such as the breakdown of the predicted time-development in accordance with the Schr\"odinger equation, the abruptness of change in a measurement ("natura non facit saltum", where art thou?),  the apparent non-applicability of quantum rules to macroscopic systems, imputed arbitrariness of Born's probability rules, the requirement of "infinite regress" for the measuring apparatus and others. Numerous papers enlarged on these  issues \cite{Leggett, GPW2} and various proposals for resolution of the problem were put forward. These include the observer's cognition \cite {Wigner}, stochastic effects \cite{Louisell}, in particular spontaneous localization \cite {GPW2, GhirardiRW1,  Pearle},  a many world scenario \cite{Everett}, non-linearity addition  to the Schr\"odinger equation \cite{BBM}, Poincar\'e recurrent state \cite {Sewell}, gravitationally induced collapse \cite{Diosi, Penrose, Donadietal}, etc.

 ~
 Common to these works, and with the specific purpose of providing a blue-print for measurements compatible with   the Copenhagen formulation of quantum theory, was the need to give expression to the coupling of the microscopic system with its macroscopic  environment. Standing apart from these and  belonging to the field of non-equilibrium thermodynamics and to the establishment of equilibrium, a general form for  this interaction was given  by  Lindblad     \cite {Lindblad} and by Gorin , Kossakowski and Sudarshan \cite{GoriniKS}, satisfying some necessary conditions. Constructing a merger between  the two separately oriented fields, S. Weinberg recently  proposed a Lindblad-operator mechanism for the collapse of the density matrix (DM) in the course of a complete measurement \cite {Weinberg}. Notably, the mechanism  was linear in the state's DM. The collapsed state (Eq. (1) in \cite {Weinberg}) comprises the set of projection operators of the measurable item; the system's Hamiltonian is described by a spectral decomposition onto the same operators (Eq. (16) in \cite {Weinberg}) (although in the verbal discussion a more general situation is considered): collapse is achieved "independent[ly] of the details of these [Lindblad] operators". Decay between energy eigenstates had earlier been treated by the Lindblad formalism (for a pedagogical presentation the volume \cite {NielsenC}, Chapter 8 may be consulted) employing the interaction representation. However, this is not convenient for treating measurements of observables that do not commute with the Hamiltonian. Detailed theories relate to the outcome ("mapping") of quantum operations, including measurements; the present work describes the process of these happening.(For a pedagogical introduction to stochasticity-induced wave-packet-  reduction, obviating pointer reading, one may refer to \cite{wpreduc}.)
 \section {Overview of the Method and Terms}\label{Overview}

 ~

 \subsection {The leading idea, also in review}\label{leading}

 ~

  While the concept  of unity of  observer  and observation had already featured in Bohr's view: "The answer that we get is built up from the combined interaction of [the observer's] state and the object of interrogation." \cite{Bohr}, this was not given a formal expression in the Copenhagen interpretation. It was more emphatically asserted both by J. Bell: "I meant that the 'apparatus' should not be severed from the rest of the world in boxes ...\cite{Bell}" and A. Peres: " A measurement  both  creates  and records a property of the system \cite {Peres}". This change in the course of a measurement affects also the environment outside the observed system ; in the words of A. Leggett
   "...under these conditions the macroscopic apparatus, and more generally
any part of the macro-world which has suffered changes in the course of the
measurement process, does not end up in a state with definite macroscopic
properties at all,... \cite{Leggett}".

~

 The same line of thought appears to motivate  S. Weinberg, who wrote in his preamble to a 2016 Lindbladian formulation of the masurement process\cite{Weinberg}, that "We will instead  [of the original formulation  of the Copenhagen interpretation, (which we will not dwell on here)] adopt the popular \textit{modern } view that the Copenhagen interpretation refers to open systems in which the transition is \textit{driven}
by the ineraction of the microscopic system under study (which may include an observer) \textit{chosen } to bring the transition about." (Our italics.)

~

These developments indicate the justification for a formulation in which the effect of the apparatus is incorporated in the equation defining the evolution of the system, rather than one in which the two entities are separate, barring an interaction between them.

  \subsubsection {"Alignment"}

   The process whereby the pointer readings become in correspondence with the possible values of the observable. Formally, for $I$ possible values, the combined density matrix reduces from comprising $I^2$ terms to one having $I$ terms. (E.g., equation (2.5) in \cite {Leggett}.)

   \subsubsection {"Dissipator"}

Added term (in the form of sums of appropriately weighted jump-operators) to the standard time dependent Schr\"{o}dinger equation, inducing non-unitary evolution in the system, accompanied by changes of its information entropy.

 \subsection {Motivation for the choice of formalism}

 ~

 Thermalization  of open systems can be described by a Lindbladian
 formalism in which Gibbsian probabilities are so inserted as parameters, that the "Dissipator" vanishes at these values of the density matrix. Replacement of the Gibbsian probabilities by Born probabilities achieves alignment in a state reduction and does so continuously.

 ~

 Limitations: Born's probability rules are assumed, not derived; the interaction term is not traced to a microscopic mechanism.

 ~

The source of this interaction term, shown in Eqn. \ref{Lop} below, incorporating the coupling  between the observed system and its surroundings  (including the measuring device) is an open question (also raised by a referee). In its application to a thermalization process, the Lindbladian jump operators have been derived, though with the aids of several  approximations (e.g.,\cite {BreuerP}), as well as, more recently,  for the dissipation in a Dicke system with a bosonic background \cite {Jageretal}. We are not in the position to provide such first principle derivation for the Lindbladian jump-operators bringing about a transition and incorporating the Born rules. It seems to be specific to the type of measurement under consideration and it is clear that just \textbf{any} jump operator, as in Weinberg's Lindbladian formulation will not do the job . Likely, one would need to include non-Markovian dynamics, so that the coupling to the device and eventual
pointer reading are \textbf{two} separate consecutive events. Inclusion of such dynamics is outside the scope of  the present work.

\section{Assumptions} \label{assumptions}
We explore the time (t)-development of the \textit{combined} density matrix $\rho(t)$  of the measured system (S) and  of the reading (pointer, dial, etc.)  on the measuring apparatus (A) for a complete and discrete measurement , expressing the underlying assumptions by three propositions.

 Proposition 1. In accord with the long-time historical approach, the measured object S and the pointer of the measuring set-up  A are treated on equal footings as subject to microscopic quantum laws, and formally describable by their respective Hamiltonians. Aware of the difficulties connected with an "infinite regress", the effects of the rest of the Universe on  S+A are not  included in the formalism; instead, for a phenomenological, approximative description, a Lindbladian term appears in the master equation.

Proposition 2. Prior to the measurement with A and S decoupled, and being free of external influence for a long time, both are in energy quantum states, pure or mixed.  After the measurement, the state is not an energy eigenstate and subsequently it will spread over to a superposition of energy eigenstates. The fast decoherence case treated below in section \ref{fast} is akin to the Zeno effect \cite {Zeno}.

Proposition  3.   Only those states of the reading apparatus (e.g., the right or left positions of a pointer) that may be in  direct correspondence with the measured states of the system (e.g., spin up or down) are given expression in the formalism. (At a beginning, the case treated is one in which there is a one-to-one correspondence between the states of the system and the readings of the apparatus; a generalization is given subsequently.) A  discussion in section~ 8 touches on  the epistemological status of the Lindbladian terms in a measurement process.

 \section{Analysis}\label{analysis}
 Considering (for simplicity) a pure state for the system, its initial state-vector written in the basis of the observed property $|S,i>$ takes the form
\beq \psi^S(t=0)=\sum_{i=1,..,I} c^S_i|S,i>\label{psiS}\enq
Born's rule for the probability of observing the $i$-component is $|c^S_i]^2\equiv p_i$, summing to unity.  Likewise, for the apparatus readings $j$, numbering $J$, one has the superposition with (complex and normalized) coefficients $c_j^A $  \beq \psi^A(t=0)=\sum_{j=1,..,J} c_j^A|A,j>\label{psiA}\enq
We start with the one-to-one correspondence situation, for which $I=J$, and the reading $j$ on A establishes uniquely the value $i=j$ for the system's measured property.

For the  combined state-vector the density operator has the outer-product form (where the stars denote complex conjugates):
  \beq\sum_{i,j,i',j'} |S,i>|A,j>c^{S*}_ic^{A*}_j c^A_{j'} c^S_{i'} <A,j'|<S,i'|\equiv\sum_{i,j,i',j'} |i,j>C_{iji'j'} <i',j'|\label{RSA1}\enq the right hand side written in an obvious shortened notation, in which $C_{iji'j'}=c^{S*}_ic^{A*}_j c^A_{j'} c^S_{i'}$.
After collapse, the density operator takes the aligned, single-sum form
 \beq \sum_{i}|S,i>|A,i>|c^{S}_i|^2<S,i|<A,i|\label{Rfin}\enq

  It will be now shown that this is the time-asymptotic solution of the Lindbladian master equation properly parametrized.

We recall Lindblad's equation for the time
 varying density of states operator $\rho\equiv \rho(t)$, as being of the  following general form:
\beq
\frac{\partial\rho}{\partial t}  = -\frac{i}{\hbar}[H,\rho]+\sum_{n}\gamma_n\langle L_n\rho L_n^{\dagger}
 -\frac{1}{2}(L_n^{\dagger}L_n\rho + \rho L_n^{\dagger}L_n)\rangle
 \label{L0}
 \enq
The second term, here named the "Lindblad term" \cite{Lindblad, GoriniKS} though in different contexts also referred to as  the Dissipator \cite{FunoSS},  contains $L_n$'s  that are  Lindblad jump-operators. We shall consistently work in the observable $+$ pointer's basis (i.e., \textit{not} in an energy basis). In this basis, neither the density operator $\rho=\rho(t)$, nor the A+S Hamiltonian $H$ is diagonal at the beginning or in the course of the development. But, as will be demonstrated, the Lindbladian formalism, by a proper choice of its form, drives  A+S to the desired  diagonal form for the combined observable $+$pointer basis. We  postulate just one single term in the previous n-sum, as well as  off-diagonal  forms, namely $|i,j><i',j'|,~ ( i,j \neq i',j')$, for the jump-operators  in the observable basis, leading to the following parametrized form of the Lindblad term
\ber
{\cal L} \rho &\equiv& \Gamma \Omega
\sum_{i',j' \neq i,j  }\frac{r(i,j)}{r(i',j')}\langle |i,j><i',j'|\rho|i',j'><i,j|
\nonumber \\
 &-& \frac{1}{2}(|i',j'><i,j|i,j><i',j'|\rho + \rho|i',j'><i,j|i,j><i',j'|\rangle
\label{Lop}
\enr

%\ber\{\script{L}\rho\} &\equiv &\Gamma\Omega \sum_{i,j\neq i',j'}\frac{r(i,j)}{r(i',j')}\langle}|i,j><i',j'|\rho|i,j><i',j'|\nonumber\\
%&- &\frac{1}{2}(|i',j'><i,j|i,j><i',j'|\rho + \rho|i',j'><i,j|i,j><i',j'|\rangle \label{Lop1}\enr

   Here a circular frequency $\Omega$  is inserted, so as to make $\Gamma$ , that quantifies the strength of the system-environment coupling, dimensionless. One notes that in the pre-factor appear the parameters $r(i,j),r(i',j') (i,j,i',j'=1,...,I)$ whose significance will be clear by deriving the matrix elements of the above operator. These are

   \ber
   {\cal L}\rho_{i,j,i',j'} &=&    \delta_{i,i'}\delta_{j,j'}r(i,j)
   \sum_{k,l}r^{-1}(k,l)\rho_{k,l,k,l}
   \nonumber \\
   &-&\frac{1}{2}[r^{-1}(i,j)+r^{-1}(i',j')]\rho_{i,j,i',j'}\sum_{k,l} r(k,l)
   \label{Lmatrix}
   \enr
   It can be seen that the trace of the above vanishes and that each matrix element vanishes upon the substitution \beq \rho_{i,j,i',j'}= \delta_{i,i'}\delta_{j,j'}r^2(i,j)\label{rho1}\enq
   While these properties hold for any arbitrary $r(ij)$, the observable-pointer alignment is achieved by  identifying  the $r$ parameters with the system's  superposition coefficient: $r(ij)=|c^S_i|\delta_{i,j}$, or
   \beq r(i,j)^2=|c^S_i|^2 \equiv p_i\delta_{i,j}\label{rp}\enq
    the last being the Born probabilities appearing in the collapsed state. As already noted, this identification  of probabilities relates to the well known procedure for the Lindblad-induced thermalization  of open  systems, for which detailed balance imposes the relation between the pre-factors $\gamma(\delta E)$/$\gamma(-\delta E)=e^{-\beta\delta E}/Z$, the latter being the canonical probabilities (with $\beta=1/k_BT$, $k_B$ the Boltzmann constant, $T$ the ambient temperature and $Z$ the partition function \cite{FunoSS, HorowitzP, BreuerP}) .

    \
 ~

 [It also seems fair to point out that also in  the standard (Copenhagen, or von Neumannian) description of the alignment stage, as appears in e.g. Eq.(2.5) of \cite {Leggett}, this development is summarily stated, without specification of the underlying mechanism.]

   \section {Fast Decoherence Limit}\label{fast}
We now consider the case that the time development in the state is predominantly due to the coupling to the environment, rather than to the unitary change induced by the Hamiltonian, meaning that the second term on the right hand side in Eq. \ref{L0} dominates the first. Quantitatively: $\Gamma>>||H||/\hbar\Omega$. Neglecting  the commutator we now form matrix elements of the Lindblad term in Eq. \ref{L0} in the observable+pointer basis. Because of the approximation made, the off-diagonal matrix elements are decoupled from the diagonal ones. The master equation of the off-diagonal terms reads (with a notation simplified by writing for the index pairs $i,j\rightarrow r$, $ i',j'\rightarrow s$ and consequently for
$\rho_{i,j,i',j'} \rightarrow \rho_{rs}\equiv\rho_{rs}(t)$
\beq
\frac{d\rho_{rs}}{dt} =
-\Gamma\Omega
\left[ \frac{\sqrt{p_r}+\sqrt{p_s}}{2}\rho_{rs}\sum_m \sqrt{p_m} \right],~r\ne s
\label{Rrs}
\enq
This shows that off-diagonal matrix elements decay exponentially in time (decohere), maintaining their real character that they had initially. Had we kept the (imaginary) commutator term, we would have found that the decay is  modulated by the eigen-energies of the Hamiltonian.

For the diagonal matrix elements we find,
\beq
\frac{d\rho_{rr}}{dt}=\Gamma\Omega
\left[ \sqrt{p_r}\sum_m\frac{\rho_{mm}}{\sqrt{p_m}}-
\frac{\rho_{rr}}{\sqrt {p_r}}\sum_m\sqrt{p_m}\right] \label{Rrr}
\enq

Again, it can be seen that the trace of the last expression  vanishes, and so does the right-hand side  under the substitution $\rho_{rr}\rightarrow p_r$. With these taking the values as in Eq. \ref{rp}, one arrives at the aligned form (written out in the  original, system-pointer indexes)
 \beq \rho(t\rightarrow\infty)=\sum_i |\psi^A_i>|\psi^S_i>|c^S_i|^2<\psi^S_i|<\psi^A_i|\label{Rend}\enq

\subsection{Illustrative example for a two-way experiment}\label{example}
Exemplifying the foregoing  for a two-valued system (such as a $\frac{1}{2}$-spin electron), prepared
as an eigenstate of a Zeeman-field with the magnetic field inclined  at an angle $2\alpha^S$  to  the vertical, in conjunction with an apparatus pointer, represented as being likewise in an eigenstate of a quasi-Zeeman field inclined at an angle $2\alpha^A$  to the vertical. The eigenstates are linear superpositions  of their $z$- spins; these are the observables that  are  to be determined by the measurement. Initially, the system and the pointer are in the superposition states as shown above in Eqs. \ref{psiS} and \ref{psiA} and whose  superposition coefficients $c^S_i$ and $c^A_j$  now  have the values,  $\sin/\cos (\alpha^S)$ and $\sin/\cos (\alpha^A)$, respectively. The DM in the observable basis  is now a $4$x$4$ matrix, in which appear all the combinations of the products of the above circular functions. As the outcome of the application of the Lindblad operator in the rate equation, at long times  the matrix   becomes reduced to the diagonal form discussed  earlier. In these, $\cos^2 (\alpha^S)=p_{1}$ and $\sin^2 (\alpha^S)=p_{4}$  belonging to the aligned observable  lie on the diagonal and are non-zero; the other two diagonal entries for the anti-aligned situations are zero.

Plotted in Figure 1 are computed DM  eigenvalues as functions of time (in red and blue), normalized to their respective Born probabilities, showing their asymptotic convergence. In  green, the typical decohering tendency of an off-diagonal element is demonstrated. Figure 2 depicts  the entropy $S(t)=-\sum_rP_r(t)\log P_r(t)$ of the system and apparatus-pointer, (in which $P_r(t)$ are computed eigenvalues of the DM.) The non-monotonic behavior is characteristic of of the Lindblad formalism, in which the environment's entropy change is not taken into account.

[In numerical work, based on forward integration, putting zeros for some of the $p_i$'s introduces singularities, eventually algebraically cancelling out, but preventing flow of computation. Therefore, instead, one puts  arbitrarily small values for these and obtains for the aligned DM  one that is arbitrarily close to, but not exactly equal to the true one.]

\begin{figure}
\includegraphics[width=12cm,height=6cm]{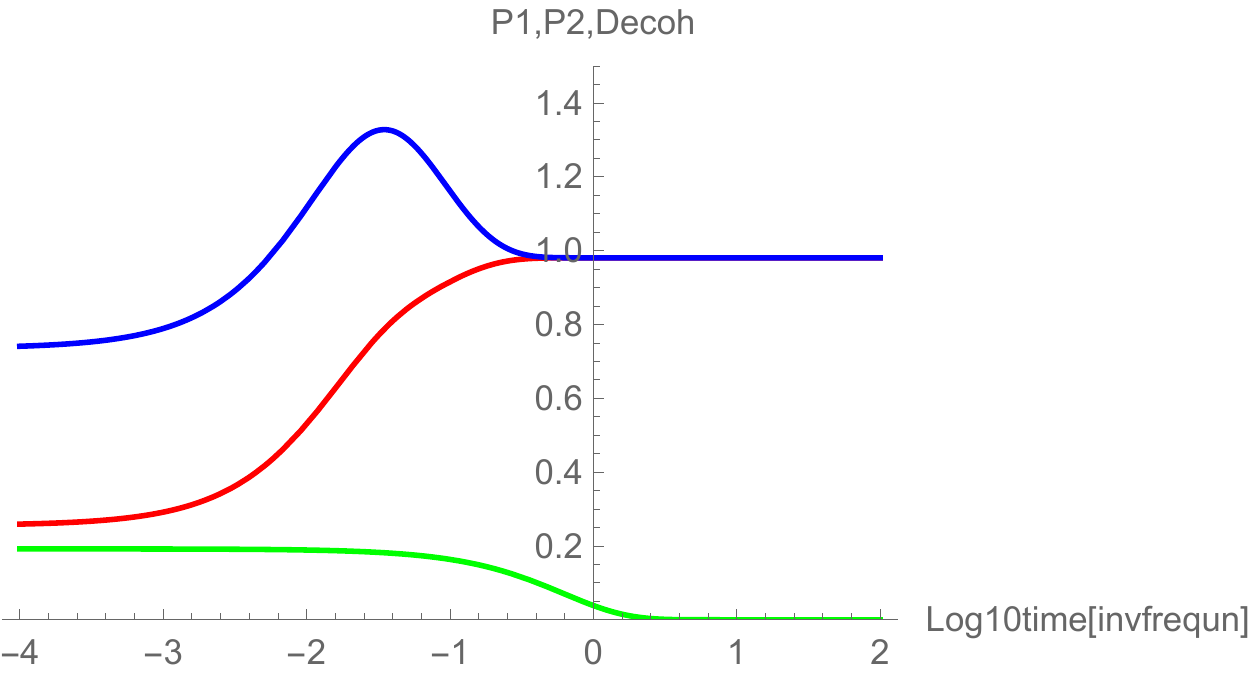}
\caption {Density matrix eigenvalues normalized to their asymptotic (pointer-aligned) values for the two aligned terms in the illustrative example (in red and blue), plotted against time in inverse circular frequency unit. In green is shown a decohering off-diagonal matrix element. Lindblad coupling strength $\Gamma=5$, $\alpha$ angles $.37~\pi$ and $.65~\pi$.}
\label{QMp1p4a}
\end{figure}

 \section{Eigenvalue analysis}\label{eigenvalue}
An alternative to the numerical solution  of the differential rate equation is eigenvalue analysis, already treated in \cite{Weinberg}, based on the Landbladian term being  a linear function of the diagonals in the density matrix. Thereby, the resulting   rate equations have  solution of the form \beq \rho_{nn}(t)= \sum_{k}v_{n,k}e^{\lambda_k t}\label{v}\enq in which $\lambda_k$ and $v_{n,k}$ are the diagonalized eigenvalues and eigenvectors of the Lindbladian matrix diagonals in Eq. \ref{Rrr}. Calculation shows that for the $4$ x $4$ matrix considered above there are three negative eigenvalues and one zero eigenvalue, which alone is of interest at the long term behavior. Belonging to this eigenvalue, the (transposed) eigenvector is found to be $\{p_1,p_2, p_3, p_4\}\approx\{c^S_1,0,0,c^S_4\}$, as  required for the alignment between the quantum states and the reading in the measuring apparatus.
\begin{figure}
\includegraphics[width=12cm,height=6cm]{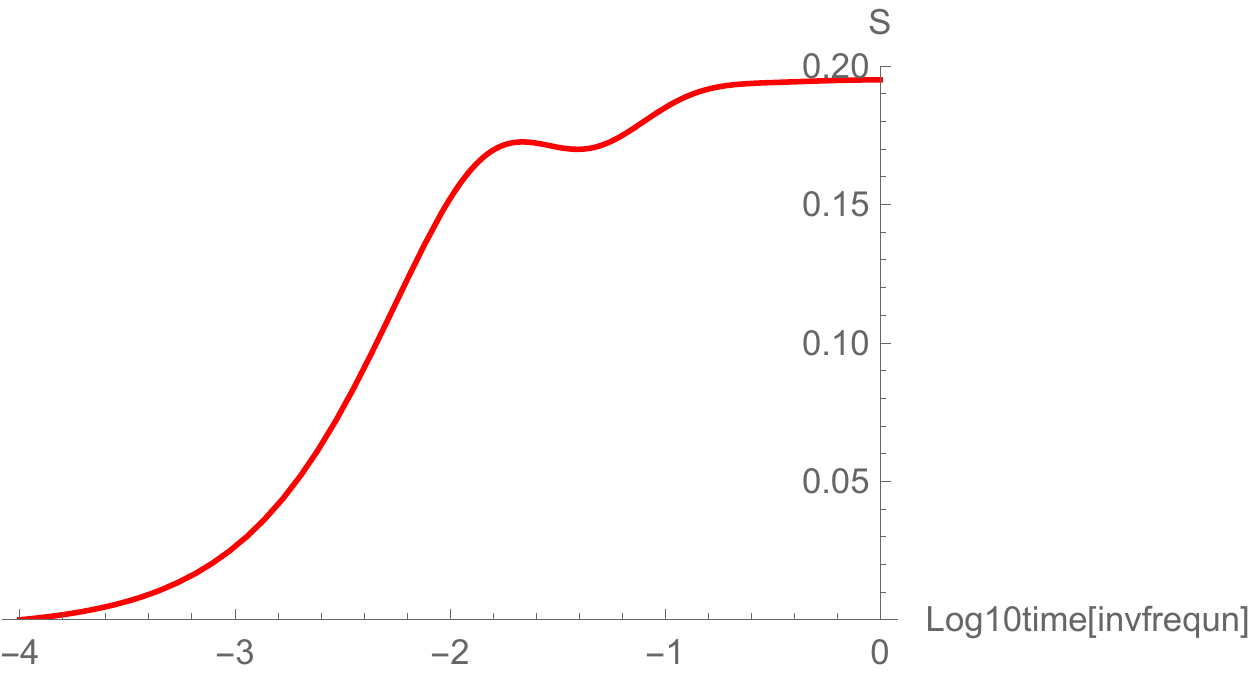}
\caption{Entropy of the combined system \textit{plus} apparatus. Noteworthy is the initial peak common to the Lindblad formalism .}
\label{QMEnt}
\end{figure}

\subsection {Measurement speed}
\label {speed}

Figure 1 shows that alignment is achieved for the model with the chosen strength parameter ($\Gamma=5$) by a time of  cca. $0.1/\Omega$. By varying the strength in the computed model, we find a  shortening  of this time that is inversely proportional to the strength. This is expected from the quantum speed limit (QSL) results that border quantum transition times $\tau$ from below.

Essentially, QSL is the ratio of two norms \cite{Mandelshtam, MargolusL}, that of the "quantum distance" \cite {BraunsteinC} and of the speed of the state evolution. Formally \beq \tau >\frac{||\rho(t\rightarrow\infty)-\rho(t=0)||}{||\frac{d\rho(t)}{dt}||}=\frac{||\rho(t\rightarrow\infty)-\rho(t=0)||}{||[L\rho(t)]||}\label{QSL1}\enq Ways of calculating the norms vary, e.g., \cite{DeffnerC, AnandanA}. Recently, for a system developing due to a Lindbalian operator, three contributions to the speed were discerned \cite{FunoSS}. To estimate $||\rho(t\rightarrow\infty)-\rho(t=0)|$, we have used the "Trace Distance "defined as
\beq T(\rho,\sigma)=\frac{1}{2}Tr[\sqrt(\rho-\sigma)]=\frac{1}{2}\sum_i|\mu_i|\label{dist}\enq
\cite {Barnett}, where $\mu_i$ are the eigenvalues of the matrix differences. The DM velocity, as defined above , changes (decreases) with time, ultimately vanishing at the fulfilment of alignment; we have taken the root-mean-square sum of the rate of the diagonal matrix elements at initial times. These yield a very low limit of
\beq \tau > \frac{1}{2}\frac{1.41}{5 * 69.03}=.0045/\Omega\label{QSL2}\enq
to be compared with the actually computed value, about 20 times longer. Better (higher) limits of transition times may be generated by different ways of forming the norm for the DM velocity (e.g. not at the beginning).

\subsection{Multiple Reading-System correspondence} \label{multiple}
 A simple generalization of the foregoing applies when each (eigen-)value of the observable is in correspondence with not just one reading of the pointer, but with several (say, R) readings, all of the same significance for the outcome. Then one simply inserts $p_i/R$ into the corresponding Lindblad term, in place of just $p_i$. In the more
complex case, that not all readings have the same likelihood, $p_i$ would have to be weighted by a probability factot, rather than by a constant denominator.

\section{The Lindbladian, "Who ordered this?"}\label{who}
Historically,  Lindblad terms were introduced as the most general forms that maintain complete positivity of the DM's and preserve their trace \cite{Lindblad,GoriniKS}. The various derivations   that have been presented (and among these a recent one   by \cite{Manzano}), involve several approximations  for the coupling between the system and its environment. Insomuch that the derivation involves also tracing over the degrees of freedom of the environment, much  detail of the latter is lost and of course it is impossible to work backwards from the Lindbladian to the environment. What is remarkable is that for special purposes the appropriate Lindbladian operators take a very special, practically unique form. Such is  the case for the accepted description of thermalization \cite{BreuerP, FunoSS} by a Lindblad formalism. The parametrization of the Lindblad term employed in the present work, though it may appear arbitrary and particular for each case,  is in fact identical to the one used for thermalization subject to the relabelling of the Gibbsian thermal distribution function as (the  Born) probabilities ($p_1,p_2,...$),with the proviso of working in the observable, rather than in the energy basis.  (This contrasts with the different approach in \cite {Weinberg}, which claims attainment of collapse for \textit{any} Lindbladian operator.) At the same time, it needs to be noted that the analog of detailed balance is missing in wave-function collapse. How come to have such a specific Lindbladian, whose source may be  \textbf{any} measurement device and procedure? One is left to  wonder about the possibility of a special meta-physical status of the Lindblad terms, or query with Wheeler "Who ordered this?"

\section{Conclusion}\label{conclusion}
The well-known Lindbladian extension to the quantum theory of motion to environmental effects is here adapted to establish the resolution of a wave-packet in a measurement as a smooth process. This is enabled by an unambiguous parametrization of the jump-operators describing the interaction of the broad environment with the observed system, both regarded as quantal entities.

~

  Above, in section 2.1, a brief historically oriented  preview has been provided for  the distinct approach in this work, namely, one based on the wholeness of the entities (observed system and observing device), through  the  (Lindbladian)  equation yielding the evolution of the system.

~

A main result emerging from the formalism, and capable of experimental verification,  is the finitely temporal variation of the system, and this in a deterministic way rather than just statistically, on the average, contrasting also with the instantaneous collapse description by (e.g.) von Neumann. Such temporal variation in continuous-thermalization processes has been proposed quite recently \cite{KorzekwaL1, KorzekwaL2}, also  by employment of  a Lindbladian formalism and within a Markovian framework.

Experimentally, verification of the time dependence of the transition in any particular measurement, implicit in our formulae, could be observed by repeated observation performed on the system subject to non-demolition transitions. These observations would be akin to the Zeno-effect measurement, such as has been achieved in the form of quasi-periodic oscillation of the result for a Superconducting  flux cubit\cite {Kakuetal}. Further work is needed for quantifying the information-entropy change in the environment \cite{Barnett, Keyl}.
\vspace{0.7 cm} \\
\begin{Large}
{\bf Acknowledgement}
\end{Large}

~

~ The authors thank the referee for meticulous reading and insightful questioning of a previous version of this paper.

\end{document}